\documentclass[amsfonts]{article}
\usepackage{graphicx}
\pdfoutput=1
\usepackage{subcaption}

\newcommand{\be}{\begin{equation}}
\newcommand{\ee}{\end{equation}}
\newcommand{\bea}{\begin{array}}
\newcommand{\ea}{\end{array}}
\newcommand{\beqa}{\begin{eqnarray}}
\newcommand{\eeqa}{\end{eqnarray}}

\newcommand{\eean}{\end{eqnarray*}}

\def\up#1{\leavevmode \raise.16ex\hbox{#1}}

\newcommand{\gapproxeq}{\lower
 .7ex\hbox{$\;\stackrel{\textstyle >}{\sim}\;$}}
\newcommand{\lapproxeq}{\lower .7ex\hbox{$\;\stackrel
{\textstyle <}{\sim}\;$}}

\newcounter{appendice}

\def\thebibliography#1{{\bf REFERENCES\markboth
 {REFERENCES}{REFERENCES}}\list
 {[\arabic{enumi}]}{\settowidth\labelwidth{[#1]}\leftmargin\labelwidth
 \advance\leftmargin\labelsep
 \usecounter{enumi}}
 \def\newblock{\hskip .11em plus .33em minus -.07em}
 \sloppy
 \sfcode`\.=1000\relax}

\begin{document}
\centerline{ \LARGE Growing  Hair  on the  extremal $BTZ$ black hole }

\vskip 2cm
\centerline{B. Harms\footnote{bharms@ua.edu} and A. Stern\footnote{astern@ua.edu} }
\vskip 1cm
\begin{center}
{ Department of Physics, University of Alabama,\\ Tuscaloosa,
Alabama 35487, USA\\}
\end{center}
\vskip 2cm
\vspace*{5mm}
\normalsize
\centerline{\bf ABSTRACT}
We show that the nonlinear $\sigma-$model  in an asymptotically  $AdS_3$ space-time admits  a  novel local   symmetry.  The field action is assumed to be quartic in the nonlinear $\sigma-$model fields and minimally coupled to gravity.  The local symmetry transformation simultaneously twists  the nonlinear $\sigma-$model fields and changes the space-time metric, and it can be used to map the extremal $BTZ$ black hole to infinitely many hairy black hole solutions.

\bigskip
\bigskip

\newpage

Asymptotically $AdS$ black holes are of current interest, in part due
to their correspondence to thermal states in the conformal theory.
Asymptotically $AdS$ black holes have been shown to admit hair.
Solutions with various types of hair have been constructed, and their
properties have been investigated
\cite{Dehghani:2001nz}-\cite{Li:2017gyc}.
The search for solutions for $AdS$ black holes with scalar hair can
require introducing complicated expressions for the potential energy
density. In the present work we are
able to generate an infinite number of asymptotically $AdS_{3} $ black
hole solutions with scalar hair. This is possible due to the presence of a hidden local symmetry, rather than some potential energy term.  

The   hair examined in this article is
associated with the nonlinear $\sigma$-model fields. Classical solutions to
the nonlinear $\sigma$-model coupled to gravity were obtained
previously, and they corresponded to topological
solitons.\cite{Harms:2016pow} No
hairy black solutions were found. The Lagrangian utilized in
\cite{Harms:2016pow} was invariant under $SO(3)$
transformations in the target space, quadratic in the $\sigma$-model
fields and minimally coupled to gravity in $2+1$ dimensions. In the
present work we replace the $\sigma$-model action in
\cite{Harms:2016pow} by one which is quartic in the nonlinear
$\sigma$-model fields (and $SO(3)$ invariant). In flat space-time there are no static solutions to this dynamical system.  On the other hand, one can find solutions   in flat space-time when symmetry breaking terms are added.\cite{Adam:2010jr}  No such symmetry breaking terms are needed for this purpose after coupling to gravity, as we show in this work.
 Using a general ansatz
for a rotating field configuration in a stationary space-time, we find
that the system of Einstein equations, field equations and $AdS$
boundary conditions determine the fields only up to an arbitrary radial
function. This indicates the existence of a local (radially dependent)
transformation which leaves the combined gravity-$\sigma$-model action
invariant. The symmetry transformation simultaneously twists  the
sigma model fields and changes the space-time metric. The
transformation can be used to map the extremal BTZ black
hole~\cite{Banados:1992wn} to infinitely many hairy black hole
solutions. Unlike  the previously found hairy black hole solutions,
here the energy-momentum tensor is traceless, and the Hawking
temperature is zero (just as with the case of the extremal black hole). These results are unchanged under the action of
the local symmetry transformation.

The outline for this article is the following: We first introduce the
nonlinear $\sigma$-model with a quartic field action minimally coupled to
gravity in $2+1$ dimensions and with a negative cosmological constant.  We
then write down an ansatz for the nonlinear $\sigma$-model fields in a rotating
space-time. The fields and metric tensor are shown to be undetermined
by the Einstein equations, field equations and $AdS$ boundary
conditions, and  this is due to the presence of a local symmetry
which simultaneously  changes the metric tensor and twists the nonlinear $\sigma$-model fields.
For different choices of an arbitrary radial function we show that we
can obtain hairy black holes, which we argue are \textit{extremal}
hairy black holes. We find naked singularity solutions, as well. We
conclude with some final remarks.

The nonlinear $\sigma$-model fields shall be denoted by $\Phi^{a}$,
$a=1,2,3$, where $\Phi^{a}\Phi^{a}=1$. The fields are maps to unit
vectors in a three-dimensional internal space i.e., the target space is
$S^{2}$. The action is generally taken to be invariant under $SO(3)$
rotations of the fields in the internal space, as will be assumed here.
Rather than applying the standard quadratic field
action coupled to gravity, as was done in \cite{Harms:2016pow}, we assume
an action for $\Phi^{a}$ which is \textit{quartic} in the fields and
their derivatives. For the full action of the nonlinear $\sigma$-model
coupled to gravity with a negative cosmological constant we take
%
\begin{eqnarray}
S&=&\int d^{3} x\sqrt{-g}\,\Bigl(\frac{1}{16\pi G}( R-2\Lambda)\;-
\frac{1}{4} F^{a}_{\mu\nu} F^{\mu\nu a}+ \lambda(\Phi^{a}\Phi^{a}-1)
\Bigr)+ S_{GHY} \;, \label{srtngactn}
\end{eqnarray}
where $G$ is the three-dimensional version of Newton's constant (here
in dimensionless units), $\Lambda<0$ is the cosmological constant and
$\lambda$ is a Lagrange multiplier. $F^{a}_{\mu\nu}$ may be interpreted as field
strengths for the nonlinear $\sigma$-model, which can be formulated as a
gauge theory.\cite{Balachandran:1978pk,Balachandran:1991zj}
They can be expressed in terms of derivatives of $\Phi^{a}$ according
to
%
\begin{eqnarray}
F_{\mu\nu}^{a}=\epsilon^{abc}\partial_{\mu} \Phi^{b}\partial_{ \nu}
\Phi^{c}\;, \label{smdlfldstrs}
\end{eqnarray}
where $\epsilon^{abc}$ is the Levi-Civita symbol. $\Phi^{a}$ and $F^{a}_{\mu\nu}$  transform as 
vectors under the action of the $SO(3)$ group, $\Phi^{a}\rightarrow
\Phi^{\prime\,a}=\mathcal{R}^{ab}\Phi^{b}$, $F_{\mu\nu}^{a}\rightarrow F_{\mu\nu}^{\prime\,a}
=\mathcal{R}^{ab}F_{\mu\nu}^{b}$ , $\mathcal{R}\in SO(3)$, and
hence $S$ is $SO(3)$ invariant. $S_{GHY}$ is the Gibbons--Hawking--York
term~\cite{York:1972sj} written on the boundary at spatial infinity
$r\rightarrow\nobreak\infty$.  We assume that the space-time is asymptotically $AdS_{3}$.  It may be necessary to add divergent counter
terms to (\ref{srtngactn}) in order for the action evaluated
on the space of solutions to be finite. We shall not consider them
here, as they do not affect the classical dynamics.

The Einstein equations and field equations resulting from variations of
the action with respect to the metric tensor $g_{\mu\nu}$ and the
fields $\Phi_{a}$ in (\ref{srtngactn}) are
\begin{eqnarray}
R_{\mu\nu}-\frac{1}{2} R g_{\mu\nu}+\Lambda g_{\mu\nu} &=&8 \pi G
\,T_{\mu\nu} \label{Instneqs} \\
&&\cr
\partial_{\mu} J^{\mu a}\;
&\propto & \Phi^{a} \label{fldeqfrFi}\;,
\end{eqnarray}
respectively, where the energy-momentum tensor $T_{\mu\nu}$ and
currents $J^{\mu a}$ are
%
\begin{eqnarray}
T_{\mu\nu} &=&F_{\mu\rho}^{a} F_{\nu}^{\;\rho a} -\frac{1}{4} g _{\mu\nu}
F^{a}_{\rho\sigma} F^{\rho\sigma a} \label{Tmunu} \\ &&\cr
J^{\mu a}& =&\sqrt{-g}\, \epsilon^{abc}\,F^{\mu\nu b} \partial_{
\nu}\Phi^{c} \label{thefldeq}\;.
\end{eqnarray}
The field equations (\ref{fldeqfrFi}) state that $J^{\mu a}$
is conserved in directions perpendicular to $\Phi^{a}$ in the target
space. From (\ref{fldeqfrFi}) one easily obtains the
conservation of the stress-energy tensor. The assumption that the
space-time is asymptotically $AdS_{3}$ means that the invariant interval
goes to
%
\begin{eqnarray}
ds^{2}\rightarrow\Lambda r^{2} dt^{2} -\frac{dr^{2}}{ \Lambda r^{2}}+
r^{2}d\phi^{2}\;, \qquad \mathrm{as} \;\; r\rightarrow\infty
\label{smptcads}\;,
\end{eqnarray}
where $\Lambda$ is again the cosmological constant. Then from
(\ref{Tmunu}), the energy density at large distances becomes
$T_{tt}\rightarrow\frac{1}{2}r^{2}( \Lambda^{2} F_{r\phi}^{2}- \Lambda
F_{rt}^{2}+\frac{1}{r^{4}} F^{2}_{t\phi})$. It is positive definite
since~$\Lambda<0$. The requirement that $T_{tt}$ vanishes in the large
distance limit means that all the field strengths vanish in the limit.
This in turn implies that the tangent vectors $\partial _{\mu}\Phi^a$ are
parallel or zero when $ r\rightarrow\infty$. From the former
possibility it follows that the $\sigma$-model fields need not point
along a constant direction in the target space at the $AdS_{3}$
boundary. This differs from the case of the $2+1$ dimensional
$\sigma$-model with the usual quadratic Lagrangian density
$\sim\partial_{\mu}\Phi^{a}\partial^{\mu}\Phi^{a}$. In that case energy
finiteness necessarily compactifies any time-slice of the space-time to
$S^{2}$, and field configurations belong to disjoint equivalence classes
associated with $\Pi_{2}(S^{2})$. While this restriction does not follow from
energy finiteness in our case, here we shall only consider field
configurations $\{\Phi^{a}(t,r,\phi)\} $ which do point along a
constant direction in the internal space at the $AdS_{3}$ boundary and
hence do belong to disjoint equivalence classes.

We restrict to stationary metrics, allowing for a nonvanishing angular
momentum. Following \cite{Harms:2016pow}, the most general such metric
can be expressed in terms of three radially dependent functions $A$,
$B$ and $\Omega$ according to
%
\begin{eqnarray}
ds^{2}=-A(r) dt^{2} +\frac{B(r)}{A(r)}{dr^{2}}+ r^{2}\Bigl(d\phi+
\Omega(r)dt\Bigr)^{2} \label{rotatngmtrc}\;.
\end{eqnarray}
Since the space-time is asymptotically $AdS_{3}$,\footnote{More
generally, $\tilde{\Omega}$ can tend to a nonzero constant. However,
the constant vanishes if we transform to the co-rotating frame.}
%
\begin{eqnarray}
A\rightarrow-\Lambda r^{2} \qquad \qquad B\rightarrow1 \qquad \qquad
\Omega\rightarrow0 \label{adslimit}\;,
\qquad \mbox{as} \;\;
r\rightarrow\infty\;. \label{smptABO}
\end{eqnarray}
For $\Omega\ne0$ the space-time has a nonvanishing angular momentum,
i.e., it is rotating. In order to obtain nontrivial solutions to
(\ref{Instneqs}) and (\ref{fldeqfrFi}) it is
necessary to assume that the nonlinear $\sigma$-model fields rotate as
well in the target space. Following \cite{Balachandran:1991zj}
rotational invariance for the nonlinear $\sigma$-model fields means
that $\partial_{\phi}\Phi ^{a}+\epsilon^{ab3}\Phi^{b}=\nobreak0$. The condition
for rotating fields is
$\partial_{t}\Phi^{a}=-\omega\partial_{\phi}\Phi^{a}$, where $\omega$ is
the angular velocity. The solution can be written in terms of one
radially dependent function $\chi$,
%
\begin{eqnarray}
\pmatrix{ \Phi^{1} \cr \Phi^{2} \cr \Phi^{3} } = \pmatrix{
\sin\chi(r)\,\cos{(\phi-\omega t)} \cr \sin\chi(r)\,\sin{(\phi-\omega
t)} \cr \cos\chi(r) } \label{rotatngfld}\;.
\end{eqnarray}
The tangent vector $\partial_{r}\Phi^{a}$ is orthogonal to $\partial
_{t}\Phi^{a}$ and $\partial_{\phi}\Phi^{a}$, and so from the previous
energy considerations it should vanish at the $AdS_{3}$ boundary. The
field strengths (\ref{smdlfldstrs}) resulting from the ansatz
are $F_{\mu\nu}^{a}=f_{\mu\nu}\Phi^{a}$, where $f_{\mu\nu}$ are independent of $t$, $f_{rt} = -\omega f
_{r\phi}=\omega\, \partial_{r}\cos\chi$ and $ f_{t\phi} = 0$, and in
this sense the field configurations (\ref{rotatngfld}) are
stationary. Using (\ref{Tmunu}) and (\ref{smptABO})
the asymptotic behavior of the energy density for these configurations
is $T_{tt} \rightarrow\frac{1}{2} \Lambda
r^{2}(\Lambda-\omega^{2})(\partial _{r}\cos\chi)^{2} $ as
$r\rightarrow\infty$. The requirement of finite energy means that
$\cos\chi$ goes asymptotically to a constant. Then in fact all tangent
vectors $\partial_{\mu}\Phi^{a}$ vanish at the $AdS_{3}$ boundary. We
choose $\chi\rightarrow0$ as $r\rightarrow \infty$. In order for
$\Phi^{a}$ to be well-defined at the origin, $\chi(0)$ needs to be an
integer multiple of $\pi$, and so field configurations
(\ref{rotatngfld}) belong to $\Pi_{2}(S^{2})$. Such
configurations can yield an everywhere well-defined energy density
$T_{tt}$ even in the presence of a space-time singularity.

Upon substituting (\ref{rotatngmtrc}) and (\ref{rotatngfld}) into the
action (including the Gibbons--Hawking--York term) one gets
%
\begin{eqnarray}
S&=&\frac{\pi}{\kappa}\int\frac{ dtdr}{\sqrt{B}} \;\biggl\{ {\partial_{r}
A}+\frac{r^{3}(\partial_{r}\Omega)^{2}}{2}+\frac{2 rB}{\ell^{2}} - {\kappa\ell^{2}}\, r\Bigl(\frac{A}{r^{2}}-{(\omega+\Omega)^{2}}
\Bigr)(\partial_{r}\chi)^{2}\sin^{2}\chi\biggr)\; \biggr\}\;,
\end{eqnarray}
where $\kappa\ell^{2}=8\pi G$, and we set $\Lambda=-\frac{1}{\ell
^{2}}$. It is convenient to introduce the dimensionless radial variable
$x=r/\ell$. Then the action becomes
%
\begin{eqnarray}
S=\frac{\pi}{\kappa}\int\frac{ dtdx}{\sqrt{B}} \;\biggl\{ { A'}+\frac{
x^{3}\tilde{\Omega}^{\prime\,2}}{2}+{2 xB}- { \kappa x}H
\chi^{\prime\,2} \sin^{2}\chi\biggr\}\;, \label{dmnslsactn}
\end{eqnarray}
where
%
\begin{eqnarray}
H=\frac{A}{x^{2}}-(\tilde{\omega}+\tilde{\Omega})^{2}\;,
\end{eqnarray}
$\tilde{\Omega}=\ell\,\Omega$ and $\tilde{\omega}=\ell\,\omega$. The
prime denotes a derivative with respect to $x$. The Einstein equations
resulting from extremizing the action with respect to variations in
$A$, $B$, $\tilde{\Omega}$ are
%
\begin{eqnarray}
\frac{x}{2} (\ln B)'& =&\kappa\,\chi^{\prime\,2}\sin^{2}\chi \label{vryA}\\
A' &=&2x B -\frac{x^{3}\tilde{\Omega}^{\prime\,2}}{2} + {\kappa\,x}H\,
\chi^{\prime\,2}\,\sin^{2}\chi \\
\left(\frac{x^{3}\tilde{\Omega}'}{\sqrt{B}}\right) ' &=& \frac{2 \kappa
x}{\sqrt{B}}\,(\tilde{\Omega}+\tilde{\omega})\,\chi^{\prime\,2}
\,\sin^{2}\chi \label{vryOmg}
\end{eqnarray}
respectively, and the field equation resulting from variations with
respect to $\chi$ is
%
\begin{eqnarray}
\biggl( \frac{ xH}{\sqrt{B}}\,\chi'\,\sin\chi\biggr)' &=&0 \;.
\label{eqsdmrnqdratic}
\end{eqnarray}
These equations can be obtained directly from
(\ref{Instneqs}) and (\ref{fldeqfrFi}).
(\ref{eqsdmrnqdratic}) is a consequence of the conservation
of $ J^{\mu a}$ in directions normal to~$\Phi^{a}$. It means that
$\frac{ xH}{\sqrt{B}}\,\chi'\sin\chi$ equals a constant, $C$. From the
asymptotic conditions (\ref{adslimit}),
$(\cos\chi)'\rightarrow-\frac{C}{x}$ as $x\rightarrow\infty$. Since
$\cos\chi$ cannot be logarithmically divergent in the limit, the
constant $C$ must be zero. For nontrivial matter fields, this further
means that $H$ must vanish for all $x$, and so
%
\begin{eqnarray}
A= {x^{2}}(\tilde{\Omega}+\tilde{\omega})^{2}\;.
\label{xsqD}
\end{eqnarray}
$A$ is then a positive function, and furthermore
(\ref{eqsdmrnqdratic}) cannot be used to determine the matter
degree of freedom $\chi$. The angular velocity $\tilde{\omega}$
appearing in the ansatz for the nonlinear $\sigma$-model fields is not
arbitrary. In order for (\ref{xsqD}) to be consistent with
the $AdS_{3}$ limit at spatial infinity (\ref{adslimit}), we
need that $\tilde{\omega}^{2}=1$. We choose $\tilde{\omega}=1$. Using
(\ref{xsqD}), the remaining equations
(\ref{vryA}-\ref{vryOmg}) reduce to just two
independent equations
%
\begin{eqnarray}
\frac{x}{2} (\ln B)' &=&\kappa\,\chi^{\prime\,2}\sin^{2}\chi \label{fftn}\\
\Bigl(x^{2}(\tilde{\Omega}+1)\Bigr)' &=&{2x}\sqrt{B} \label{sxten}\;.
\end{eqnarray}
There are only three relations,
(\ref{xsqD})-(\ref{sxten}) amongst the four
functions $A$, $B$, $\tilde{\Omega}$ and $\chi$, and so solutions are
parametrized by an arbitrary function. We can choose, for example, the
arbitrary function to be the matter field $\chi$, and then use it to
determine the space-time metric. Alternatively, the space-time metric
can be expressed in terms of an arbitrary function, say
$\tilde{\Omega}$, from which one can then determine the matter field.

The results can be compactly summarized after making the coordinate
change to a frame which is co-rotating with the nonlinear
$\sigma$-model fields: $(t,x,\phi)\rightarrow(t,y,\xi)$, where
$y=x^{2}({\tilde{\Omega}+1})$ and $\xi=\phi-t$. While $x$ is defined on
the half-line, the range of $y$ depends on the function
$\tilde{\Omega}(x)$. [From (\ref{sxten}), the coordinate
change is valid globally provided that $B$ does not vanish at some
$x\ne0$.] Using $\tilde{\omega}=1$, the ansatz
(\ref{rotatngfld}) for the $\sigma$-model fields depends only
on $y$ and $\xi$, $\Phi^{a}=\Phi ^{a}(y,\xi)$. In the $(t,y,\xi)$
coordinates, the metric tensor is given by
%
\begin{eqnarray}
ds^{2}=\frac{dy^{2}}{4y^{2}}\,+\, x(y)^{2} d\xi^{2}\,+\,2{y}\,d\xi dt
\;, \label{mtrcntyxi}
\end{eqnarray}
where here $x$ is regarded as a function of $y$. The $AdS_{3}$ vacuum
has $ x(y)^{2}=y$ and  $\chi(y)$=constant. More generally, $ x(y)^{2}$ is related to $\chi(y)$
by
%
\begin{eqnarray}
\frac{d^{2}}{dy^{2}}\,x(y)^{2}\,+\,2\kappa\,\Bigl( \frac{d}{dy}
\cos\chi(y)\Bigr)^{2}\,=\,0\;, \label{Instinxiy}
\end{eqnarray}
which is the only condition that results from the Einstein equations.
The only nonvanishing components of the field strength in these
coordinates are $F^{a}_{\xi y}=(\frac{d}{dy}\cos\chi)\, \Phi^{a}$, and
the only nonvanishing components of the current $ J^{\mu a}$ are $ J^{t
a}=-2y\, \partial_{\xi} \Phi^{a}$. Since this current is identically
conserved, no conditions on $\chi(y)$ result from the field equations
(\ref{thefldeq}).

The presence of infinitely many solutions shows the existence of a
continuous symmetry of the action. This is evident from the result that
the integrand of the action evaluated on the space of solutions is
independent of $x(y)^{2}$ or $\chi(y)$. In fact it is a
constant.\footnote{The action evaluated on the space of solutions can
be made to vanish by adding a divergent term to
(\ref{srtngactn}).} The infinitesimal version of the symmetry
transformation is
\begin{eqnarray}x(y)^{2} \rightarrow x(y)^{2}+\delta x(y)^{2}\;,\qquad
\chi(y)\rightarrow \chi(y)+\delta\chi(y)\;,\label{locsym}
\end{eqnarray}
where
$\frac{d^{2}}{dy^{2}}\,\delta x(y)^{2}=-4 \kappa(
\frac{d}{dy}\cos\chi(y))\frac{d}{dy}(\delta\cos\chi(y))$. It
simultaneously changes the $g_{\xi\xi}$ component of the metric, and
twists the nonlinear $\sigma$-model fields, $\Phi(y,\xi)
\rightarrow\Phi(y,\xi)+\delta\Phi(y,\xi)$, where
%
\begin{eqnarray}
\delta\Phi(y,\xi)&=& R_{3}(\xi)\,T_{2}\, R_{3}(\xi)^{-1}\,\Phi(y,
\xi)\,\delta\chi(y)\; ,\end{eqnarray}
\be R_{3}(\xi)= \pmatrix{ \cos{\xi}&-\sin{\xi}&
\cr \sin{\xi}&\cos{\xi}& \cr &&1 } \quad \qquad T_{2}= \pmatrix{ &&1
\cr && \cr -1&& } \;.\ee

The system may admit an event horizon. $y=$ constant hypersurfaces are
space-like or null. This is because the normal one-form $dy$ has norm
equal to $g^{yy}=4y^{2}\ge0$. So a $y=0$ slice, if it exists, is
everywhere null and thus an event horizon. $K=\partial_{t}$ [or
equivalently, $K=\partial_{t}+\partial_{\phi}$ in $(t,x,\phi)$
coordinates] is a null Killing vector on the event horizon, meaning
that it is also a Killing horizon. In fact, $K=\partial_{t}$ is null on
the entire space-time manifold. Furthermore, $K^{\mu} \nabla_{\mu} K^{
\nu}$ is identically zero implying that a Killing horizon has zero
surface gravity. From this one concludes that an event horizon would
have zero Hawking temperature. This result holds for all possible
matter contributions consistent with (\ref{rotatngfld}),
including no matter contribution.  The result is thus invariant under the 
local symmetry transformation  (\ref{locsym}). For the case of no matter contribution,
the system reduces to the extremal $BTZ$ black hole (or the $AdS_{3}$
vacuum), which we show next.

The simplest postulate for the nonlinear $\sigma$-model fields is
$\Phi^{a}=\delta^{a3}$, or $\chi=0$, yielding a vanishing
energy-momentum tensor. Then from (\ref{Instinxiy}) and the
$AdS_{3}$ boundary condition one gets that $ x(y)^{2}=y+
{M_{\infty}}/{2}$, where $M_{\infty}$ is an integration constant. The
resulting range for $y$ is: $- {M_{\infty}}/{2}\le y<\infty$. While
$M_{\infty}=0$ gives the $AdS_{3}$ vacuum, the $M_{\infty}\ne0 $
solution describes an extremal BTZ black hole. This is obvious when
re-expressed in the $(t,r,\phi)$ coordinates. The functions $A$, $B$
and $\tilde{\Omega}$ for the solution are
%
\begin{eqnarray}
A= x^{2} -M_{\infty} + \frac{M_{\infty}^{2}}{4x^{2}}
\qquad
\qquad
B=1
\qquad
\qquad
\tilde{\Omega}=-\frac{ M_{\infty}}{2x^{2}}\;,
\label{xtrmlbh}
\end{eqnarray}
which agrees with the standard expression for a BTZ black
hole~\cite{Banados:1992wn}. Because $M_{\infty}$ corresponds to
both the mass \textit{and} the angular momentum, the black hole is
extremal, with
a single horizon at $x=x_{0}=\sqrt{M_{\infty}/2}$,.

More generally, $\chi$ is an arbitrary function satisfying
$\chi\rightarrow0$ at spatial infinity. For the metric tensor one can
set $ x(y)^{2}=y+ M(y) /{2} $ and impose that $
M(y)\rightarrow{M_{\infty}}$ as $y\rightarrow\infty$. When $M(y)$ is
not a constant, the space-time is not locally $AdS_{3}$, meaning that
the Riemann tensor $R_{\mu\nu\rho\sigma}$ does not reduce to
$\Lambda(g_{\mu \rho}g_{\nu\sigma}- g_{\mu\sigma}g_{\nu\rho})$. In the
$(t,y,\xi)$ coordinates, the only contribution to the energy-momentum
tensor is $T_{\xi\xi}=\frac{y^{2}}{\kappa} \frac{d^{2}}{dy^{2}} M(y)$,
using (\ref{Instinxiy}). It follows that the trace of $T$ is
zero (more generally, $\mathrm{tr}\,T^{n}=0$, for any positive integer $n$).

As stated above, $\chi$ at $x=0$ is an integer multiple of $\pi$ for
our field configurations (\ref{rotatngfld}), the integer
being the winding number. (From previous arguments, energy finiteness
does not require that the winding number be conserved.) Below we
examine a winding number one field configuration. It is easiest to
express it in $(t,r,\phi)$ coordinates. Take
%
\begin{eqnarray}
\chi(x)=2 \tan^{-1}\frac{a}{x}\;,
\label{sterographic}
\end{eqnarray}
where $a$ is a scale factor. Then $[\Phi^{a}]^{-1}$ defines the
stereographic projection of the unit sphere to the plane. From
(\ref{xsqD})-(\ref{sxten}) it follows that
%
\begin{eqnarray}
B=e^{ -2\rho}\;, \qquad \qquad
\rho=\frac{8}{3} \frac{\kappa a^{4}}{(x^{2}+a^{2})^{3}}\nonumber\\
y =x^{2}(\tilde{\Omega}+1)\;=\;\frac{1}{3}(x^{2}+{a^{2}}{})\, E_{
\frac{4}{3}} (\rho) \;+\; C{} \label{twntee3} \;,
\end{eqnarray}
where here $E_{n}$ denotes the exponential integral $E_{n}(\rho)=
\int_{1}^{\infty} {dt}\, t^{-n}e^{-\rho t}$, and $C$ is an integration
constant. The result (\ref{twntee3}) follows from the
identity $(n-1)E_{n}(\rho)=e^{-\rho}-\rho E_{n-1}(\rho)$. To obtain the
asymptotic expansion for $A$ and $\tilde{\Omega}$ one can use
$E_{\frac{4}{3}} (\rho)= 3 + \Gamma[-\frac{1}{3}]\rho^{1/3}+ \mathcal
{O}(\rho)$. In comparing the asymptotic form of $A$ and
$\tilde{\Omega}$ with (\ref{xtrmlbh}) one then gets
%
\begin{eqnarray}
-\frac{ M_{\infty}}{2}=(8\kappa)^{\frac{1}{3}}\Bigl(\frac{a}{3}
\Bigr)^{\frac{4}{3}}\Gamma[-\frac{1}{3}]+a+C\;. \label{Minfty}
\end{eqnarray}
$B$ approaches a  finite nonzero value as $x$ tends to zero, while $A$
and $\tilde{\Omega}$ diverge as $\sim1/x^{2}$ for generic values of
$C$. This leads to a singularity in the metric tensor at $x=0$.
Nevertheless, the energy density is well behaved near the origin. From
$\chi\sim\pi-\frac{2x}{a}$, one gets that the energy density $
T_{tt}=\frac{A}{B}\chi^{\prime\,2}\sin^{2}\chi$ has a finite limit as
$x\rightarrow0$.

$A$ and $\tilde{\Omega}$ are nonsingular at the origin for the special
case

\noindent
 $i)$ $ C=C_{0}$,
%
\begin{eqnarray}
C_{0}=-\frac{a^{2}}{3}\, E_{\frac{4}{3}}(\frac{8\kappa}{3a^{2}})\;.
\label{czero}
\end{eqnarray}
For this case, $A$ vanishes at the origin, $A\rightarrow0$ and
$\tilde{\Omega}+1\rightarrow\frac{1}{3}\, E_{\frac{4}{3}} (\frac{8
\kappa}{3a^{2}})$ as $x\rightarrow0$. A~vanishing $A(x)$ for some value of $x$
indicates the existence of a horizon. Since $ C=C_{0}$ means that $A(x)$ vanishes at the origin, this
corresponds to the limiting case of a horizon at the origin. $A$ as
a function of $x$ is plotted for this case in {figure one} [solid
curve]. $\kappa$ and the asymptotic value of the mass $ M_{\infty}$
were set to one in {the figure}. $a$ and $C_{0}$ can then be determined
from (\ref{Minfty}) and (\ref{czero}),
respectively. The result is $a\approx0.7264$ and
$C_{0}\approx-.00018$.

\begin{figure}[h]
\centering
  \includegraphics[height=2.5in,width=3.5in,angle=0]{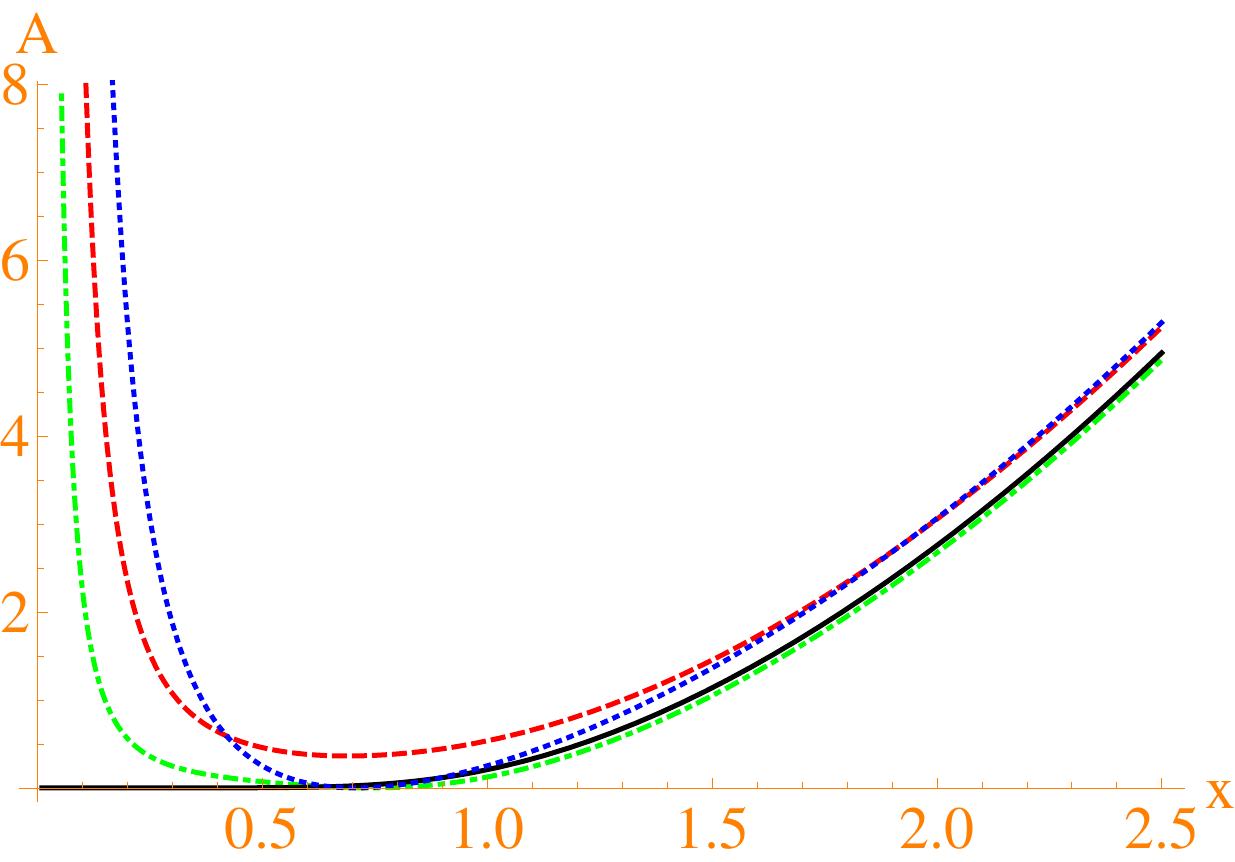}
\caption {$A$ versus $x$ for  a fixed value of the asymptotic mass and the coupling, $ M_\infty=\kappa=1,$ and various values of $C$ and $a$ consistent with (\ref{Minfty}).  Horizons occur at zeros of $A$. 
 Case i)  [ solid curve] has $C=C_0=-.00018$ and  $a\approx 0.7264$, and it is the limiting case of a horizon at the origin. 
  Case ii) [dot-dashed curve] has $C=-.15<C_0\approx 0$ and  $a\approx  .5998$, and exhibits a horizon at
 $x=x_0\approx .7406  $. 
Case iii) [dashed curve] has $C=.3>C_0\approx -.038$ and $a\approx  .9475$  with no horizon.  The plots are compared to $A$ versus $x$ for the extremal BTZ black hole [dotted curve] with $ M_\infty=1$ and horizon at  $\approx .7071$. }  
\label{fig:test}
\end{figure}

Aside from this limiting case, there exist two other classes of
solutions which follow from (\ref{sterographic}){:}

\noindent
 $ii)$ For $C<C_{0}$, $A$, and also $\tilde{\Omega}+1$, have a
    zero at some finite nonzero value $x_{0}$ of $x$, indicating an
    event horizon at $x=x_{0}$. The configuration corresponds to a
    hairy black hole. Even though it has a nonvanishing source, it
    shares properties with the extremal black hole, e.g. it has
    only one horizon (despite the possibility of having
    nonvanishing angular momentum), $g^{yy}\ge0$, and the Hawking
    temperature is zero. We can then say that this is an
    {\it extremal} hairy black hole. In {figure one}~$A$ as a
    function of $x$ is plotted for $C=-.15$ [dot-dashed curve].
    Again we have set $\kappa= M_{\infty}=1$, while $a$ is
    determined from (\ref{Minfty}). The result is
    $a\approx.5988$. Also from (\ref{czero}),
    $C_{0}\approx 0$, and so $C<C_{0}$. For these values of the parameters there is a
    horizon at $x=x_{0}\approx.7406 $. This can be compared to an
    extremal BTZ black hole of mass equal to one [dotted curve in
    {figure one}], which has a horizon with radius $\approx.7071$.

\noindent
 $iii)$ No horizons occur for $C>C_{0}$. Such configurations
     have a naked singularity at the origin. $A$ as a function
    of $x$ is plotted for $C=.3$ in {figure one} [dashed curve].
    After setting $\kappa= M_{ \infty}=1$, (\ref{Minfty})
    and (\ref{czero}) give $a\approx.9475$ and
    $C_{0}\approx-.038<C$, respectively.

Similar results are obtained after replacing
(\ref{sterographic}) with other ans\"{a}tse, for example,
ones with winding number greater than one.

We conclude with the following remarks:

 Whether or not the local symmetry we found plays a role in the corresponding conformal theory is an open
question. The  symmetry applies only for \textit{extremal} black
holes and \textit{extremal} hairy black holes.  No analogous local symmetry which mixes the nonlinear
$\sigma$-model and the space-time metric was found in the case of
non-extremal black holes. We also found no evidence of this symmetry
when the quadratic term in the nonlinear $\sigma$-model is included in
the action.\cite{Harms:2016pow}

It has been noted that many properties of extremal black holes are distinct
from non-extremal black holes, and that the extremal limit of
non-extremal black holes is singular. For example, their topologies
differ.\cite{Teitelboim:1994az}  Also unlike in the non-extremal case, extremal black holes have zero Hawking
temperature and zero
{entropy.\cite{Preskill:1991tb}-\cite{Das:1996rn}}
It was also found that extremal and nonextremal black holes have
distinct greybody factors and quasinormal mode
{structure.\cite{Gamboa:2000uc}-\cite{Crisostomo:2004hj}}
The extremal hairy black hole solutions we obtained here also 
have properties which are distinct from that of other hairy black holes.  Extremal hairy black holes have a traceless energy momentum tensor and zero Hawking temperature. It
would be of interest to know if there are other properties which are
unique to extremal hairy black holes.

The energy momentum tensor (\ref{thefldeq}) for our system
has the same form as for electromagnetism. It is therefore natural to
ask whether or not similar solutions exist in the $2+1$
Einstein--Maxwell system. Given  analogous expressions for the
electromagnetic field strengths $f_{\mu\nu}$ in a stationary
space-time, namely $ f_{\xi y}= f(y)$, $ f_{\xi t}= f_{y t}= 0$, one
can solve the Einstein equations for the metric tensor given in
(\ref{mtrcntyxi}). Instead of (\ref{Instinxiy}),
the Einstein equations now give $\frac{d^{2}}{dy^{2}}\,x(y)^{2}+2
\kappa\,f(y)^{2}=0$. While $ F_{\xi y}$ is undetermined for the
nonlinear $\sigma$-model we have been investigating, $ f_{\xi y}$, and
hence $x(y)^{2}$, is fixed in the electromagnetic theory. From the
sourceless Maxwell equations one gets
%
\begin{eqnarray}
f_{\xi y}= \frac{q}{y} \qquad \quad x(y)^{2}=y+ \frac{M_{\infty} }{2}
+2\kappa q^{2} \,\ln y \;,
\end{eqnarray}
where $q$ and $ M_{\infty}$ are constants, and we again assume that the
space-time is asymptotically $AdS_{3}$. The general set of rotationally
invariant solutions to Einstein--Maxwell equations in $2+1$ dimensions
(which applies to the nonextremal case as well) was found
in~\cite{Cataldo:2002fh}.

Of course, it is natural to ask whether or not this system can be
generalized to higher dimensions, and specifically if a local symmetry
between nonlinear model fields and the space-time exists in that
setting. A~natural candidate in four space-time dimensions would be the
$SU(N)$ chiral model coupled to gravity. The Skyrme model coupled to
gravity in $3+1$ space-time has been studied for some time and shown to
admit solitons and hairy black
{holes.\cite{Heusler:1991xx}-\cite{Gudnason:2016kuu}}
The quadratic term in the chiral fields was present in these articles.
Our work suggests that novel results, such as the existence of a hidden
local symmetry between chiral fields and the space-time metric may
appear if the quadratic term is absent, and the only matter
contribution to the action is the quartic (Skyrme) term.

\bigskip
{\Large {\bf Acknowledgments} }

\noindent
We are very grateful to M. Kaminski, A.  Pinzul and S. Sarker for valuable discussions. 
 \end{document}